\documentclass[journal]{IEEEtran}

\usepackage{xurl}
\usepackage{cite}
\usepackage{amsmath,amssymb,amsfonts}
\usepackage{graphicx, float}
\usepackage{textcomp,enumitem}
\usepackage{xcolor}
\usepackage{multirow, booktabs, makecell}
\usepackage{algorithm2e}
\RestyleAlgo{ruled}
\usepackage{multicol}

\usepackage[ colorlinks = true,
linkcolor = blue,
urlcolor = blue,
citecolor = red,
anchorcolor = green,
]{hyperref}

\newcommand{\unitBQ}{Mvar$\cdot$s}

\def\BibTeX{{\rm B\kern-.05em{\sc i\kern-.025em b}\kern-.08em
    T\kern-.1667em\lower.7ex\hbox{E}\kern-.125emX}}

\usepackage{amsthm}
\newtheorem{property}{\textbf{Property}}
\newtheorem{proposition}{\textbf{Proposition}}

\newtheorem{theorem}{\textbf{Theorem}}

\usepackage{soul}
\sethlcolor{yellow}
\soulregister\cite7
\soulregister\ref7

\begin{document}

\title{
%Rethinking Voltage Control for Data Centers: From Nominal-Voltage Tracking to Ride-Through Awareness
Reducing Data-Center Load Transfers with Ride-Through-Aware Voltage Control
}

\author{
Siyu Lu,~\IEEEmembership{Student Member,~IEEE,}
and~Yang~Weng,~\IEEEmembership{Senior Member,~IEEE}
\thanks{
S. Lu and Y. Weng are with the School of Electrical, Computer and Energy Engineering at Arizona State University, Tempe, AZ 85281 USA, e-mail: \{siyulu12, yang.weng\}@asu.edu.}
 % \vspace{-10mm}
}

\maketitle

\begin{abstract}
Fault-induced transfers of large data centers to backup supply can cause abrupt reductions in grid-served demand. Coordinating reactive-power support to mitigate these transfers requires consideration of facility-specific voltage--duration ride-through characteristics, particularly when support capacity is limited. This paper develops a ride-through-aware extension of quadratic voltage control that assesses voltage deviations relative to the admissible margins specified by these characteristics. We establish four design requirements that jointly express envelope-based severity, duration dependence, and preventive voltage-restoration incentives. Within a multiplicatively gated quadratic class, we derive necessary and sufficient conditions for satisfying these requirements and construct an explicit family of envelope-normalized objectives. We apply the proposed objectives to reactive-power coordination and evaluate their performance on the IEEE 14-bus, New England 39-bus, and NPCC 140-bus systems. Comparisons with Volt/VAR control, conventional nominal-voltage tracking, and recent data-center voltage-control methods demonstrate improved data-center retention under matched reactive-power support constraints. In the IEEE 14-bus study, the proposed controller reduces the mean data-center disconnection rate by 44.8\% and grid-demand loss by 27.4\% compared with the best-performing baseline. These results highlight the benefit of incorporating data-center ride-through requirements into reactive-power allocation.
\end{abstract}


% ================== Introduction =============

\section{Introduction}

Large data centers are emerging as critical loads in modern power systems, posing new reliability challenges due to their high concentration of power-electronic equipment \cite{NERC_LargeLoad_Alert_2025}. Such equipment is particularly sensitive to voltage disturbances.
Grid faults can produce a voltage sag at the point of common coupling (PCC). If the sag persists beyond the equipment’s ride-through capability, the internal servers or power supplies may fall out of regulation, potentially causing server resets or shutdowns \cite{ITIC_CBEMA_Curve_2000}.
To maintain continuity of service, data centers employ uninterruptible power supply (UPS) systems that can transfer critical IT loads from grid supply to backup energy sources during such disturbances \cite{0929chen2025electricity}. However, these transfers can appear to the grid as sudden reductions in power demand. As large data centers can consume hundreds of megawatts and operate numerous parallel power conversion units, synchronized transfers may cause abrupt and substantial reductions in grid-served load \cite{NERC2025SOROverview,ERCOT2025LargeLoadStability,lu2026dynamic}.
These risks motivates coordinated voltage support to reduce avoidable such data center transfers.

\begin{figure}[h]
    \centering
    \includegraphics[width=1\linewidth]{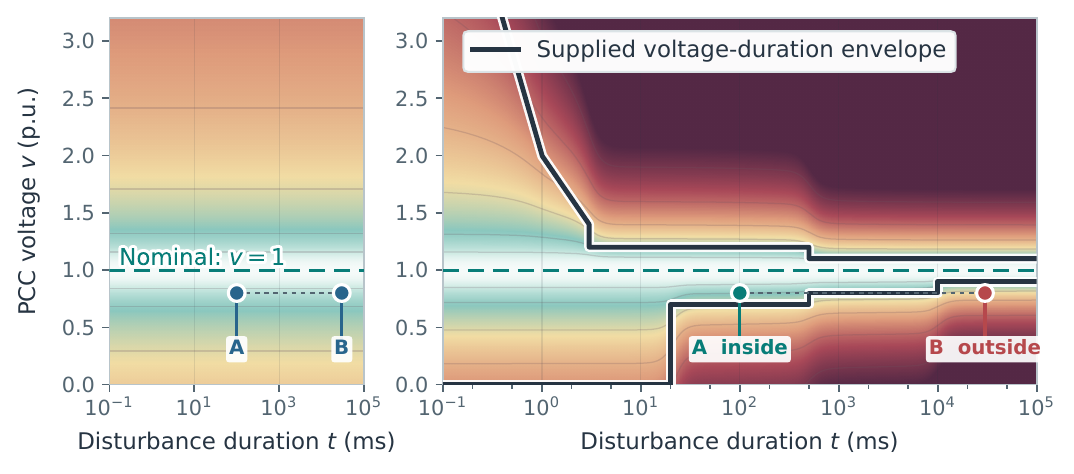}
    \caption{Comparison of conventional quadratic (left) and proposed envelope-informed (right) voltage-control objectives. Shading indicates loss magnitude. Dashed and solid lines denote nominal voltage and the ride-through envelope, respectively. At the same voltage, quadratic control equally penalizes A and B, whereas the proposed objective assigns greater severity to B for exceeding the ride-through limit.}
    \label{fig:bigpicture}
\end{figure}

Established measures for supporting voltage-sensitive loads include UPS systems \cite{AAMIR20161395}, dynamic voltage restorers \cite{Soomro2021_DVR_ComprehensiveReview}, and grid-side voltage support \cite{sarimuthu2016oltc_review,statcom}. Equipment immunity guidelines, including SEMI F47 \cite{SEMI_F47_0706} and the Information Technology Industry Council (ITIC) curve \cite{ITIC_CBEMA_Curve_2000}, characterize tolerance through disturbance magnitude and duration. As indicated by the solid boundary in the right panel of Fig.~\ref{fig:bigpicture}, a briefly tolerable voltage deviation may become unacceptable as it persists. This observation suggests that effective voltage control for data centers should account for not only the magnitude of voltage deviations but also their duration relative to ride-through limits.

For grid-side voltage support, existing methods adopt different strategies to coordinate voltage regulation.
Conventional Volt/VAR droop control offers fast local reactive-power response~\cite{narang2021ieee1547voltvar, yuan2026making}, while optimization-based methods coordinate multiple resources through network models and operational constraints~\cite{baran1989optimal,farivar2013branch,low2014convex,molzahn2017survey,lubin2019chance}. 
However, Volt/VAR droop control typically responds to instantaneous voltage measurements, whereas conventional optimization objectives primarily emphasize nominal-voltage tracking. For instance, the quadratic loss $\ell_{\rm q}(v)=(v-1)^2$ assigns identical penalties and corrective gradients to voltage disturbances of equal magnitude but different durations, as illustrated by points A and B in the left panel of Fig.~\ref{fig:bigpicture}. Consequently, these approaches do not inherently prioritize voltage support according to different data center facilities' remaining ride-through margins.

Recent studies have expanded the resources available for data-center voltage control. Building on power provisioning and workload management \cite{lin2011online}, dynamic voltage and frequency scaling enables computing demand to participate in voltage regulation \cite{chen2025voltage}. Coordinated internal resources have also been used to enhance low-voltage ride-through \cite{1004xie2025enhancing}, while online feedback optimization of distributed storage addresses voltage-limit violations and fluctuations under storage constraints \cite{Mao2026Online}. These advances and established reactive-power coordination methods \cite{zhang2015optimal,jafarian2020realtime,saxena2016cost} motivate a complementary question concerning the allocation objective. A prescribed envelope identifies admissible voltage--duration conditions, but does not uniquely determine how violations should be penalized or how restoration should be encouraged within the admissible region. In particular, a violation-only penalty has no interior voltage gradient, although a predictive controller can act before an anticipated violation. This leaves a design task: combining envelope-based severity assessment with continuous preventive voltage-restoration incentives.

Our main idea is to assess voltage deviations relative to each facility's duration-dependent tolerance and use this information to prioritize constrained control support. We formalize the intended behavior through four design requirements. The penalty should distinguish admissible conditions from violations and increase with violation depth. It should also be nondecreasing with elapsed duration at a fixed voltage under a narrowing envelope. To retain preventive restoration incentives, its voltage gradient should point toward nominal voltage and its marginal penalty should increase as a ride-through boundary is approached from within. The first two requirements express envelope alignment, while the latter two specify the corrective behavior selected for this control task. Together, they define an explicit objective-design specification rather than relying on heuristic penalty selection.

To construct an objective satisfying these requirements, we extend the conventional quadratic voltage loss through a multiplicative gating function, $\ell(v,t)=(v-1)^2g(v,t)$, where $g(v,t)$ adjusts the penalty according to voltage magnitude, disturbance duration, and the prescribed ride-through envelope. 
This formulation retains the nominal-voltage preference of conventional quadratic control while allowing the penalty magnitude and corrective gradient to adapt to ride-through requirements. 
We derive necessary and sufficient conditions on $g(v,t)$ under which the resulting objective satisfies the four requirements. 
Based on this characterization, we construct a closed-form family of compliant objectives explicitly parameterized by the ride-through envelope. This provides a systematic and analytically verifiable approach to designing voltage-control objectives for different ride-through requirements.

We evaluate the proposed method on the IEEE 14-bus, New England 39-bus, and NPCC 140-bus transmission systems. Comparisons with Volt/VAR control, conventional nominal-voltage tracking, and envelope- and duration-based baselines demonstrate improved data-center retention and reduced grid-demand loss under matched reactive-power support constraints. To assess practical applicability, sensitivity studies examine whether these improvements persist across different data-center and control-device deployments, reactive-power constraints, and fault conditions. Property-by-property ablation studies further investigate how each proposed property contributes to effective voltage regulation and data-center ride-through, while larger-system experiments assess performance and computational scalability.

The remainder of this paper is organized as follows. Section~\ref{sec:System Modeling} presents the system model and control formulation. 
Section~\ref{sec:obj_disconnection} develops the ride-through-aware objective for the control.
Section~\ref{sec:exp} presents numerical studies, and Section~\ref{sec:conclusion} concludes the paper.

% ================= System Modeling ====================

\section{System Modeling}
\label{sec:System Modeling}

We consider grid-side voltage regulation for large data centers connected to a transmission network, where fault-induced load transfers can pose significant challenges to bulk power-system stability. To reduce the risk of such transfers, the controller coordinates fast reactive-power support from devices such as static synchronous compensators (STATCOMs) to regulate voltages at data-center points of common coupling (PCCs). This section establishes the relationship between reactive-power injections and bus voltages, the operational limits of available support, and the voltage--duration information needed for control-objective design.

Let $\mathcal G=(\mathcal N,\mathcal E)$ denote a transmission network, where $\mathcal N$ and $\mathcal E$ are the sets of buses and branches, respectively. 
Data centers are connected at buses $\mathcal D\subseteq\mathcal N$, while grid-side reactive-power support, such as STATCOM, is available at $\mathcal C\subseteq\mathcal N$. 
To account for practical discrete-time control implementation, we divide the control period $T_{\mathrm{sim}}$ into $K=T_{\mathrm{sim}}/\Delta t$ control steps, indexed by $t_k=k\Delta t$, where $\Delta t$ denotes the control sampling interval.
Let $v_{i,k}=|V_i(t_k)|$ denote the voltage magnitude at bus $i$ in per unit.
Then, the vector $\Delta\mathbf q_k\in\mathbb R^{|\mathcal C|}$ collects the reactive-power injections of the controllable devices, with positive values indicating injection into the grid. We focus on reactive-power control because fast grid-side devices, such as STATCOMs, can regulate voltage without requiring active-power dispatch or access to internal data-center resources~\cite{dixon2005reactive}.

\subsection{Voltage Prediction Using Reactive-Power Sensitivities}

Accurate allocation of reactive-power support requires accounting for its network-wide voltage effects. Conventional linearized DistFlow models are primarily developed for radial distribution networks~\cite{baran1989optimal}, making them less suitable for the meshed transmission systems considered here. Instead, we adopt a voltage--reactive-power sensitivity matrix obtained by linearizing the AC power-flow equations around the corresponding operating point~\cite{zimmerman2011matpower}. The resulting voltage response is approximated as
\begin{equation}
\Delta\mathbf v \simeq \mathbf S_Q\Delta\mathbf q,
\qquad
\mathbf S_Q=\partial\mathbf v/\partial\mathbf q,
\label{eq:sq_jacobian}
\end{equation}
where $\mathbf S_Q$ captures the voltage sensitivities to reactive-power adjustments across the network. This formulation accommodates meshed transmission networks while retaining a computationally efficient linear voltage model. Fault inception and clearance can substantially alter network operating conditions and, consequently, the voltage sensitivities to reactive-power injections. To account for these changes, we update $\mathbf S_Q$ at operating-regime transitions, such as fault inception and clearance, while holding it fixed between transitions to reduce computational overhead.

At each control step $k$, the resulting voltage magnitudes are estimated using the latest operating-point information as
\begin{equation}
    \widehat{\mathbf v}_k
    =
    \mathbf v_k^0+\mathbf S_{Q,k}\Delta\mathbf q_k,
    \label{eq:voltage_prediction}
\end{equation}
where $\mathbf v_k^0$ denotes the baseline voltage magnitudes before the considered control adjustments, and $\Delta\mathbf q_k$ represents the reactive-power adjustments.

\subsection{Reactive-Power Support and Operational Constraints}

Although reactive-power adjustments $\Delta\mathbf q$ provide voltage correction, their magnitudes and response rates are constrained by the physical capabilities of supporting devices. For each device $j\in\mathcal C$, the operational constraints are given by
\begin{subequations}\label{eq:q_limits}
\begin{align}
\underline q_j
&\leq \Delta q_{j,k}
\leq \overline q_j, \label{eq:q_capacity}\\
|\Delta q_{j,k}-\Delta q_{j,k-1}|
&\leq R_j\Delta t, \label{eq:q_ramp}
\end{align}
\end{subequations}
where $\underline q_j$ and $\overline q_j$ denote the admissible reactive-power adjustment limits, and $R_j$ denotes the maximum ramp rate~\cite{IEEE1052_2018}.

Beyond individual device limits, grid operators may face restrictions on the reactive-power support allocated to data-center voltage regulation. To examine control performance under limited support availability, we consider two complementary resource budgets:
\begin{subequations}\label{eq:q_budget}
\begin{align}
\sum\nolimits_{j\in\mathcal C}|\Delta q_{j,k}|
&\le Q^{\mathrm{inst}},
\label{eq:q_budget_inst}\\
\Delta t
\sum\nolimits_{k=1}^{K}
\sum\nolimits_{j\in\mathcal C}|\Delta q_{j,k}|
&\le B_Q,
\label{eq:q_budget_total}
\end{align}
\end{subequations}
where $Q^{\mathrm{inst}}$ limits the instantaneous support available across all devices \cite{borah2026enhanced, chen2026voltage}, while $B_Q$ restricts cumulative reactive-power utilization over the control horizon \cite{du2026admittance}. These budgets allow us to investigate how different voltage-control objectives allocate limited support across data-center facilities and over time. In particular, they help assess whether incorporating ride-through requirements improves the prioritization and utilization of available reactive-power resources.

\subsection{Disturbance Duration and Objective-Design Problem}

As data-center ride-through depends on both voltage magnitude and disturbance duration, we introduce a duration state to track how long each data center experiences a voltage excursion.
For each $d\in\mathcal D$, let $\underline v_d<1$ and $\overline v_d>1$ denote the lower and upper voltage thresholds for detecting a violation. Based on the measured PCC voltage $v_{d,k}$, the violation state is defined as
\begin{equation}
s_{d,k}=
\begin{cases}
-1, & v_{d,k}<\underline v_d,\\
 1, & v_{d,k}>\overline v_d,\\
 0, & \text{otherwise}.
\end{cases}
\label{eq:disturbance_state}
\end{equation}
Then, the elapsed violation duration $\tau_{d,k}$ is updated as
\begin{equation}
\tau_{d,k+1}=
\begin{cases}
0, & s_{d,k+1}=0,\\
\tau_{d,k}+\Delta t,
  & s_{d,k+1}=s_{d,k}\ne 0,\\
\Delta t, & \text{otherwise},
\end{cases}
\label{eq:duration_update}
\end{equation}
where $\tau_{d,k}$ is initialized at zero. The duration increases while a voltage sag or swell persists and resets upon voltage recovery or a change in violation direction. At each control step, $\tau_{d,k}$ is obtained from the observed voltage history and treated as a known state in the voltage-control optimization.

With these definitions, we seek to construct a voltage-control objective that translates prescribed data-center ride-through requirements into appropriate corrective priorities. The objective-design problem is stated as follows:

\begin{itemize}[leftmargin=*]
    \item \textbf{Problem:}
    Construct a ride-through-aware voltage-control objective
    that prioritizes corrective actions according to both
    voltage deviation and disturbance duration, particularly
    under limited reactive-power support.

    \item \textbf{Given:}
    Prescribed voltage--duration ride-through requirements for each data center $d\in\mathcal D$.

    \item \textbf{Required:}
    Design a voltage- and duration-dependent loss
    $\ell_d(v,\tau)$ such that the following one-step voltage-control problem appropriately prioritizes ride-through protection:
    \begin{equation}
    \begin{aligned}
    \min_{\Delta\mathbf q_{k}}
    \quad&
    \sum_{d\in\mathcal D}
    \ell_d\!\left(
    \widehat v_{d,k},
    \tau_{d,k}
    \right)\\
    \mathrm{s.t.}\quad&
    \eqref{eq:voltage_prediction},\
    \eqref{eq:q_limits},\
    \eqref{eq:q_budget},\ 
    \eqref{eq:duration_update}, \\
    &\text{applicable network operating limits}.
    \end{aligned}
    \label{eq:generic_voltage_control}
    \end{equation}
\end{itemize}

Here, the instantaneous constraint \eqref{eq:q_budget_inst} is enforced at each control step, while the cumulative constraint \eqref{eq:q_budget_total} is considered over the entire control horizon.
Network operating constraints, including bus voltage limits and branch thermal ratings, are also considered during simulations.

% ================= Method ====================
\section{Construction of Ride-Through-Aware Voltage-Control Objectives}
\label{sec:obj_disconnection}

To appropriately prioritize voltage correction in \eqref{eq:generic_voltage_control}, the controller must distinguish data centers approaching their ride-through limits from those that can tolerate ongoing disturbances. 
The ITIC curve~\cite{ITIC_CBEMA_Curve_2000} technique provides an established reference for such assessment by characterizing equipment voltage tolerance as a function of both disturbance magnitude and duration. In particular, a voltage excursion that is initially tolerable may become unacceptable as it persists. In practice, data-center ride-through capabilities may differ due to equipment characteristics, UPS operation, and protection settings. We therefore adopt the ITIC-inspired voltage--duration envelope while allowing facility-specific ride-through requirements.

\subsection{Ride-Through Envelope and Desirable Objective Properties}
\label{subsec:ride_through_properties}

For each data center $d\in\mathcal D$, inspired by the ITIC curve technique~\cite{ITIC_CBEMA_Curve_2000} , we  represent its prescribed admissible voltage range at elapsed disturbance duration $\tau$ by
\begin{equation}
\mathcal A_d(\tau)=[a_d(\tau),\,b_d(\tau)],
\label{eq:ride_through_envelope}
\end{equation}
where $a_d(\tau)<1<b_d(\tau)$ denote the lower and upper voltage bounds, as illustrated by the solid boundary in the right panel of Fig.~\ref{fig:bigpicture}.
These duration-dependent bounds are distinct from the fixed thresholds $\underline v_d$ and $\overline v_d$ introduced in Section~\ref{sec:System Modeling}, which only initiate and reset the disturbance timer. 
Using fixed thresholds ensures that $\tau$ measures the continuous duration of a voltage excursion, independently of the ride-through limits that evolve with time.
The envelope instead specifies how the admissible voltage range changes as that timer advances. We further assume that $a_d(\tau)$ is nondecreasing and $b_d(\tau)$ is nonincreasing over the modeled duration range.

A voltage-control objective should serve two complementary purposes: reflecting the ride-through severity of voltage disturbances and providing effective guidance for corrective control. Accordingly, we establish four desirable properties to formalize these requirements. For clarity, we consider a representative data center and omit its bus index. Let $\ell(v,\tau)$ denote its voltage-control loss, and let $[a(\tau),b(\tau)]$ be the prescribed admissible voltage interval, with $a(\tau)<1<b(\tau)$. We require $\ell(v,\tau)\ge 0$ and $\ell(1,\tau)=0$.

\begin{itemize}[leftmargin=*]

\item \textit{Ride-through alignment:}
The first requirement is that the loss should distinguish voltage conditions according to their ride-through acceptability. In particular, disturbances within the prescribed envelope should receive smaller penalties, whereas violations should incur increasingly large penalties as their depths increase.

\begin{property}[P1: Envelope separation and exterior severity]
\label{prop:P1}
For a prescribed $\varepsilon>0$ and every $\tau$,
\begin{equation}
\begin{cases}
0\le\ell(v,\tau)<\varepsilon,
    &a(\tau)<v<b(\tau),\\
\ell(v,\tau)=\varepsilon,
    &v=a(\tau)\ \text{or}\ v=b(\tau),\\
\ell(v,\tau)>\varepsilon,
    &v<a(\tau)\ \text{or}\ v>b(\tau).
\end{cases}
\label{eq:P1_separation}
\end{equation}
Moreover, for $0\le h_1<h_2$ within the voltage domain,
\begin{subequations}\label{eq:P1_exterior}
\begin{align}
\ell(a(\tau)-h_2,\tau)
&>\ell(a(\tau)-h_1,\tau),\\
\ell(b(\tau)+h_2,\tau)
&>\ell(b(\tau)+h_1,\tau).
\end{align}
\end{subequations}
\end{property}

Beyond voltage magnitude, ride-through severity also depends on disturbance duration. A voltage sag that is initially tolerable may become unacceptable as it persists. The corresponding loss should therefore not decrease when disturbance duration increases.

\begin{property}[P2: Duration monotonicity]
\label{prop:P2}
For any fixed voltage $v$ and
$0\le\tau_1\le\tau_2$,
\begin{equation}
\ell(v,\tau_2)\ge\ell(v,\tau_1).
\label{eq:P2_duration}
\end{equation}
\end{property}

Together, Properties~\ref{prop:P1} and~\ref{prop:P2} establish the desired relationship between the loss and the prescribed voltage--duration envelope, including across its piecewise breakpoints.

\item  \textit{Effective corrective guidance:}
Ride-through alignment alone does not ensure that the objective is suitable for voltage-control optimization. For example, a loss that is constant throughout the admissible region cannot encourage preventive voltage correction before a boundary is reached. We therefore require the loss to provide a consistent corrective direction toward nominal voltage.

\begin{property}[P3: Informative corrective gradient]
\label{prop:P3}
At every non-nominal voltage in the design domain,
\begin{equation}
(v-1)\ell_v(v,\tau)>0,
\qquad v\ne1,
\label{eq:P3_gradient}
\end{equation}
where $\ell_v=\partial\ell/\partial v$.
\end{property}

In addition to the corrective direction, the magnitude of the gradient influences how strongly the controller responds to a voltage deviation. As an admissible voltage approaches a ride-through boundary, its remaining voltage margin shrinks, motivating a stronger marginal penalty to encourage correction before the boundary is crossed.

\begin{property}[P4: Increasing marginal penalty near the envelope]
\label{prop:P4}
Choose $\delta>0$ such that the intervals
$(a(\tau),a(\tau)+\delta)$ and
$(b(\tau)-\delta,b(\tau))$ lie within the admissible region on their respective sides of nominal voltage. Define the voltage margin $m$ as
$
m=
\begin{cases}
v-a(\tau), & a(\tau)<v<a(\tau)+\delta,\\
b(\tau)-v, & b(\tau)-\delta<v<b(\tau).
\end{cases}
$
In these neighborhoods, require
\begin{equation}
\left.
\frac{\partial|\ell_v(v,\tau)|}{\partial m}
\right|_{\tau}<0.
\label{eq:P4_margin}
\end{equation}
\end{property}

Properties~\ref{prop:P3} and~\ref{prop:P4} thus ensure that the objective provides corrective incentives within and beyond the admissible region, with stronger marginal penalties near its boundaries. These are control-design requirements rather than conditions imposed directly by the ride-through envelope.

\end{itemize}

The four properties jointly specify the desired behavior of a ride-through-aware voltage-control objective. We next investigate how they can be satisfied through a multiplicatively gated extension of conventional quadratic voltage control.

\subsection{Necessary and Sufficient Conditions for Gated Objectives}
\label{subsec:gate_characterization}

The preceding properties specify how a voltage-control loss should reflect ride-through limits and guide corrective action, but they do not prescribe its functional form. We start from the conventional quadratic voltage-deviation penalty and introduce a multiplicative gate,
\begin{equation}
\ell(v,\tau)=(v-1)^2g(v,\tau),
\label{eq:gated_loss}
\end{equation} where $g(v,\tau)$ adjusts the penalty according to the voltage magnitude and elapsed disturbance duration. This representation retains nominal voltage as a zero-loss reference while allowing both the loss and its voltage gradient to respond to the prescribed ride-through envelope. The objective is not to select the gate heuristically, but to identify precisely which gates satisfy P1--P4.

For the characterization, we assume that $g(\cdot,\tau)$ is continuously differentiable for $v\ne1$ and twice differentiable near the admissible voltage boundaries, and that \(\ell(\cdot,\tau)\) is continuous at \(v=1\).
To accommodate piecewise ride-through envelopes, we allow the gate to change across duration breakpoints while preserving the global duration monotonicity required by P2. Let
$
\mathcal U_{\delta}(\tau)
=(a(\tau),a(\tau)+\delta)
\cup(b(\tau)-\delta,b(\tau))
$
denote the interior boundary neighborhoods specified in P4. For compactness, define
$
\Gamma_g(v,\tau)
=2g(v,\tau)+(v-1)g_v(v,\tau),
$ and
$
\Lambda_g(v,\tau)
=2g(v,\tau)+4(v-1)g_v(v,\tau)
 +(v-1)^2g_{vv}(v,\tau),
$
where subscripts $v$ indicate voltage derivatives. These quantities appear directly in the voltage gradient and curvature of the loss:
\begin{equation}
\ell_v=(v-1)\Gamma_g,
\qquad \ell_{vv}=\Lambda_g.
\label{eq:loss_gradient_curvature}
\end{equation}

\begin{theorem}[Characterization of admissible gating functions]
\label{thm:gate_characterization}
Under the envelope assumptions of Section~\ref{subsec:ride_through_properties} and the regularity conditions above, the gated loss in \eqref{eq:gated_loss} satisfies P1--P4 if and only if that for every $\tau$:
\begin{subequations}\label{eq:gate_iff_conditions}
\begin{align}
g(a(\tau),\tau)
&=\frac{\varepsilon}{[1-a(\tau)]^2},
\ g(b(\tau),\tau)
=\frac{\varepsilon}{[b(\tau)-1]^2},
\label{eq:gate_boundary}\\
g(v,\tau_2)&\ge g(v,\tau_1),
\quad \tau_2\ge\tau_1,\ v\ne1,
\label{eq:gate_time}\\
\Gamma_g(v,\tau)&>0,
\quad v \neq 1,
\label{eq:gate_direction}\\
\Lambda_g(v,\tau)&>0,
\quad v\in\mathcal U_{\delta}(\tau).
\label{eq:gate_boundary_gradient}
\end{align}
\end{subequations}
The duration condition in \eqref{eq:gate_time} holds for all pairs $0\le\tau_1\le\tau_2$, including pairs across piecewise-envelope breakpoints.
\end{theorem}

\begin{proof}
For necessity, P1 requires $\ell(a(\tau),\tau)= \ell(b(\tau),\tau)=\varepsilon$; substituting \eqref{eq:gated_loss} gives \eqref{eq:gate_boundary}. For any $v\ne1$, the positive factor $(v-1)^2$ makes P2 equivalent to \eqref{eq:gate_time}; at $v=1$, $\ell(1,\tau)=0$ for every $\tau$. From \eqref{eq:loss_gradient_curvature}, $(v-1)\ell_v=(v-1)^2\Gamma_g$, so P3 is equivalent to \eqref{eq:gate_direction}. Finally, P3 establishes $\ell_v<0$ below nominal voltage and $\ell_v>0$ above it. Holding $\tau$ fixed, $m=v-a(\tau)$ at the lower boundary and $m=b(\tau)-v$ at the upper boundary. In either neighborhood,
$
\left.\frac{\partial |\ell_v|}{\partial m}\right|_{\tau}
=-\ell_{vv}=-\Lambda_g.
$
Thus P4 is equivalent to \eqref{eq:gate_boundary_gradient}.

Conversely, \eqref{eq:gate_direction} and \eqref{eq:loss_gradient_curvature} imply that $\ell$ strictly decreases toward $v=1$ from below and strictly increases away from $v=1$ above nominal voltage. Since $\ell(1,\tau)=0$, the loss is nonnegative. The boundary values in \eqref{eq:gate_boundary} therefore place all interior voltages strictly below $\varepsilon$ and all exterior voltages strictly above $\varepsilon$, with strict ordering as the violation depth increases. This establishes P1. Condition \eqref{eq:gate_time} gives P2, while \eqref{eq:gate_direction} gives P3, and \eqref{eq:gate_boundary_gradient} establishes P4.
\end{proof}

Theorem \ref{thm:gate_characterization} separates four roles of the gate. Boundary calibration aligns the loss threshold with the prescribed ride-through limits. Duration monotonicity prevents the same voltage excursion from becoming less severe as time elapses. The first-derivative condition preserves a nominal-seeking corrective direction. The curvature condition strengthens the marginal penalty when the admissible voltage margin shrinks. Notably, P1's exterior severity ordering follows directly from the boundary calibration and P3, without requiring additional conditions on the gate.

The global inequality in \eqref{eq:gate_time}, rather than only $g_{\tau}\ge0$, is intentional: it also covers abrupt changes in piecewise-defined ride-through limits. The conditions characterize a class of objectives, not a unique gate. Moreover, a gate depending only on duration generally cannot satisfy both boundary-calibration conditions when the admissible undervoltage and overvoltage deviations are unequal. We therefore construct a voltage- and duration-dependent family that accounts for this asymmetry explicitly.

\subsection{Construction of a Compliant Objective Family}
\label{subsec:objective_family}

Theorem~\ref{thm:gate_characterization} establishes the conditions on $g(v,\tau)$, but does not specify how to choose a gate for a given ride-through envelope. For grid-side voltage regulation, a useful construction should accommodate different undervoltage and overvoltage tolerances, retain a simple expression for online optimization, and allow the strength of corrective incentives to be adjusted. We develop such a family by normalizing the voltage deviation by the corresponding admissible margin.

For a representative data center, define the lower and upper admissible deviations from nominal voltage as \begin{equation} r_-(\tau)=1-a(\tau),\qquad r_+(\tau)=b(\tau)-1. \label{eq:admissible_margins} \end{equation} Both margins are strictly positive and nonincreasing with duration under the envelope assumptions of Section~\ref{subsec:ride_through_properties}. Thus, a smaller margin indicates that less voltage deviation can be tolerated on the corresponding side of nominal voltage. For any $p\ge4$, we construct the gate
\begin{equation}
 g_p(v,\tau)=\varepsilon
 \begin{cases}
 |v-1|^{p-2} / r_-(\tau)^p ,&v\le1,\\
 |v-1|^{p-2} / r_+(\tau)^p ,&v>1.
 \end{cases}
 \label{eq:constructed_gate}
\end{equation}
and define the corresponding loss $\ell_p(v,\tau)=(v-1)^2g_p(v,\tau)$. The voltage-side normalization is important: it calibrates the loss to the two ride-through boundaries even when their distances from nominal voltage are unequal. The parameter $p$ controls how strongly the loss is concentrated near and outside these boundaries.

\begin{proposition}[Compliance of the constructed family]
\label{prop:constructed_family}
Suppose $a(\tau)<1<b(\tau)$, $a(\tau)$ is nondecreasing, and $b(\tau)$ is nonincreasing, with both $r_-(\tau)$ and $r_+(\tau)$ uniformly bounded away from zero. Then, for every $\varepsilon>0$ and $p\ge4$, the gate in \eqref{eq:constructed_gate} satisfies all the conditions of Theorem~\ref{thm:gate_characterization}. Consequently, $\ell_p$ satisfies P1--P4.
\end{proposition}

\begin{proof}
At the lower and upper boundaries, $|v-1|=r_-(\tau)$ and $|v-1|=r_+(\tau)$, respectively. Hence, $\ell_p(a(\tau),\tau)=\ell_p(b(\tau),\tau)=\varepsilon$, which gives the boundary-calibration condition \eqref{eq:gate_boundary}. 
For a fixed $v\ne1$, the applicable margin $r_\pm(\tau)$ is nonincreasing in $\tau$; therefore, $g_p(v,\tau)$ is nondecreasing in duration, including across piecewise-envelope breakpoints, proving \eqref{eq:gate_time}. On either side of nominal voltage, let $r(\tau)$ denote the corresponding margin. Direct differentiation for $v\ne1$ gives 
$
\Gamma_{g_p}(v,\tau) =\frac{\varepsilon p|v-1|^{p-2}}{r(\tau)^p}>0, 
$ and 
$\Lambda_{g_p}(v,\tau) =\frac{\varepsilon p(p-1)|v-1|^{p-2}} {r(\tau)^p}>0. 
$
These establish \eqref{eq:gate_direction} and \eqref{eq:gate_boundary_gradient}. Finally, $p\ge4$ ensures that $g_p$ and its first voltage derivative are continuous at $v=1$, whereas the second derivative exists in the boundary neighborhoods. All conditions of Theorem~\ref{thm:gate_characterization} therefore hold.
\end{proof}

The construction has two implications for voltage-support allocation. First, the same loss level $\varepsilon$ is attained at either ride-through boundary, despite possible asymmetry in voltage tolerance. For a fixed normalized deviation $\rho=|v-1|/r_\pm(\tau)$, the loss is $\varepsilon\rho^p$: increasing $p$ reduces penalties for $0<\rho<1$ and amplifies them for $\rho>1$. This permits stronger emphasis on envelope-threatening deviations without eliminating the corrective gradient inside the admissible region. Second, the voltage gradient magnitude at either boundary is
$
|\ell_{p,v}(a(\tau),\tau)|=\frac{\varepsilon p}{r_-(\tau)}
$
and
$
|\ell_{p,v}(b(\tau),\tau)|=\frac{\varepsilon p}{r_+(\tau)}.
$
Accordingly, a narrower ride-through margin produces a stronger marginal penalty at its boundary. The result characterizes the objective's incentives; the resulting reactive-power dispatch and its resource efficiency still depend on the network sensitivities and operational constraints in Section~\ref{sec:System Modeling}.

For multiple data centers, \eqref{eq:constructed_gate} is applied using each facility's prescribed $a_d(\tau)$ and $b_d(\tau)$, yielding $\ell_{p,d}(v,\tau)$. The numerical studies examine how the resulting ride-through-based prioritization performs under limited instantaneous and cumulative reactive-power support.

\section{Numerical Studies}
\label{sec:exp}

Numerical experiments are conducted on the IEEE 14-bus, New England 39-bus, and NPCC 140-bus transmission systems. The experiments assess whether the proposed voltage--duration objective (i) better retains data-center load under limited reactive-power support, (ii) maintains its effectiveness and computational efficiency across different operating conditions and network sizes, and (iii) benefits from each proposed property through property-wise ablation studies.

\subsection{Experimental Setup and Comparison Methods}
\label{sec:numerical_setup}

In each test system, data centers are randomly assigned to $|\mathcal D|$ load buses (PQ buses) and assigned normalized demand traces from the MIT Supercloud dataset~\cite{MIT-data}, while the remaining load buses follow profiles from the ACTIVSg2000 dataset \cite{birchfield2016grid}. 
Network dynamics are simulated using ANDES~\cite{ANDES} over $T_{\mathrm{sim}}=0.5$~s, with a control interval of $\Delta t=1$~ms and a numerical integration step of $0.5$~ms in ANDES. 
Non-data-center loads are held constant over the short simulation period, while data-center demand is updated every $10$~ms using sampled workload traces. 

To simulate fault-induced voltage disturbances, a three-phase bus fault with reactance $x_f$, specified in Table~\ref{tab:nr_parameters}, is applied at a randomly selected bus at $t=0.05$~s and cleared at $t=0.15$~s, yielding a $100$-ms fault duration.
The sensitivity matrix $\mathbf S_{Q,k}$ in~\eqref{eq:sq_jacobian} is estimated through central finite differences around the corresponding operating condition, updated following fault inception and clearance, and held fixed between updates.

For reactive-power support, $|\mathcal C|$ buses are randomly selected to host controllable devices, such as STATCOMs, which need not be colocated with data centers. All controllers share the capacity and ramp-rate constraints in~\eqref{eq:q_limits} and the support budgets in~\eqref{eq:q_budget}. The instantaneous budget limits aggregate reactive-power adjustments at each control step, while cumulative budget $B_Q$ limits the total reactive-power utilization over the simulation period. To focus on the objective-design problem without introducing additional uncertainty from forecasting future data-center demand, we adopt an offline trajectory-normalization procedure, following the principle of aggregate-constrained profile scaling used in power-system studies~\cite{lazzeroni2023load}. Specifically, reactive-power commands are first obtained through successive one-step optimizations in \eqref{eq:generic_voltage_control} and then uniformly scaled to satisfy $B_Q$. This normalization preserves their relative temporal allocation while ensuring a consistent utilization limit for comparing different control objectives.

To represent heterogeneous data-center protection settings, the lower and upper voltage-violation thresholds $\underline v_d$ and $\overline v_d$ are independently sampled from uniform distributions over $[0.75,0.92]$ and $[1.22,1.30]$~p.u., respectively, with trip delays uniformly sampled from $[10,40]$~ms. These settings are modeling assumptions motivated by reported uncertainties in large-load ride-through and protection characteristics~\cite{ERCOT2025LargeLoadStability,NERC_LargeLoad_2026}. Each facility's violation duration $\tau_{d,k}$ is updated according to~\eqref{eq:duration_update}, and the facility transfers to backup supply when the corresponding voltage excursion persists beyond its assigned trip delay. Reconnection of data centers is not considered due to the short simulation horizon. The ITIC-inspired envelopes serve as voltage–duration severity references for preventive control rather than exact representations of equipment protection. Actual data-center transfers are determined separately by facility-specific voltage thresholds and trip delays, distinguishing continuous control guidance from discrete protection actions. Table~\ref{tab:nr_parameters} summarizes the settings across all test systems.

\begin{table}[h]
\centering
\caption{Simulation settings for the three transmission test systems.}
\label{tab:nr_parameters}
\small
\setlength{\tabcolsep}{2.5pt}
\begin{tabular}{lccc}
\toprule
Parameter & IEEE 14-bus & New England 39-bus & NPCC 140-bus \\
\midrule
$|\mathcal D|$ & $2$ & $4$ & $4$ \\
$|\mathcal C|$ & $2$ & $4$ & $4$ \\
\midrule
$[\underline q_j,\overline q_j]$ (Mvar)
& $[-100,100]$ & $[-400,400]$ & $[-400,400]$ \\
$Q^{\mathrm{inst}}$ (Mvar)
& $150$ & $600$ & $600$ \\
$R_j$ (Mvar/ms)
& $10$ & $20$ & $20$ \\
$B_Q$ (\unitBQ)
& $20$ & $60$ & $60$ \\
\midrule
$T_{\mathrm{sim}}$ (ms)
& $500$ & $500$ & $500$ \\
$\Delta t$ (ms)
& $1$ & $1$ & $1$ \\
$\Delta t_{\mathrm{ANDES}}$ (ms)
& $0.5$ & $0.5$ & $0.5$ \\
$(t_f^{\mathrm{on}},t_f^{\mathrm{off}})$ (ms)
& $(50,150)$ & $(50,150)$ & $(50,150)$ \\
$x_f$ (p.u.)
& $0.2$ & $0.02$ & $0.02$ \\
\midrule

$\underline v_d$ (p.u.)
    & $\mathcal U(0.75,0.92)$
    & $\mathcal U(0.75,0.92)$
    & $\mathcal U(0.75,0.92)$ \\
$\overline v_d$ (p.u.)
    & $\mathcal U(1.22,1.30)$
    & $\mathcal U(1.22,1.30)$
    & $\mathcal U(1.22,1.30)$ \\
$\tau_d^{\mathrm{trip}}$ (ms)
    & $\mathcal U(10,40)$
    & $\mathcal U(10,40)$
    & $\mathcal U(10,40)$ \\

$p$ & $4$ & $4$ & $4$ \\
\bottomrule
\end{tabular}
\end{table}

\begin{figure*}[t]
\centering
\includegraphics[width=1\linewidth]{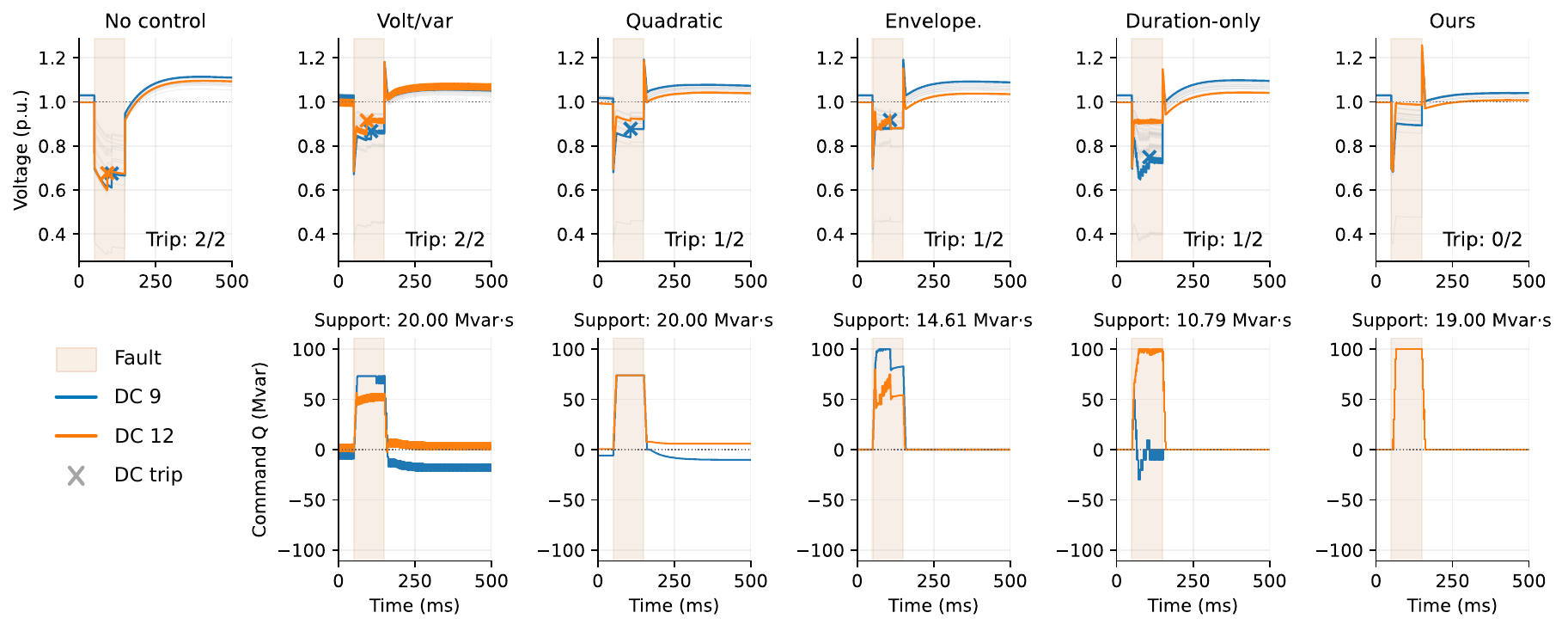}
\caption{Voltage-control comparison on IEEE 14-bus system under a cumulative control-effort budget of $20$~\unitBQ. }
\label{fig:nr_main_comparison}
\end{figure*}

We compare the proposed controller against an uncontrolled reference, two conventional voltage-regulation methods, an envelope-constrained ride-through approach, and a duration-only variant:

\begin{itemize}[leftmargin=*]
\item \textbf{\emph{No control}}: No additional reactive-power support, providing the reference data-center transfer outcomes under identical disturbances and workload profiles.

\item \textbf{\emph{Volt/VAR}}~\cite{narang2021ieee1547voltvar}: A local piecewise-linear voltage-to-reactive-power characteristic with a central deadband, representing conventional autonomous voltage regulation.

\item \textbf{\emph{Quadratic}}~\cite{1004xie2025enhancing}:
Inspired by the centralized data-center voltage controller, we implement a reactive-power-only, grid-side variant. We retain its quadratic voltage-deviation loss, $\sum_{d\in\mathcal D}(\widehat v_{d,k}-1)^2$, while excluding internal data-center resource coordination and applying the same network and reactive-power constraints as ours. This baseline represents conventional voltage regulation without explicitly considering disturbance duration or ride-through boundaries.

\item \textbf{\emph{Envelope-constrained}}~\cite{wang2026synthesizing}:
Inspired by ride-through constraint-based control developed for ERCOT's proposed voltage--duration requirements, we implement a grid-side reactive-power variant. We replace hard envelope constraints with a squared violation penalty, $\sum_{d\in\mathcal D}\big([a_d(\tau_{d,k})-\widehat v_{d,k}]_+^2+[\widehat v_{d,k}-b_d(\tau_{d,k})]_+^2\big)$, where $[x]_+=\max\{x,0\}$. This baseline explicitly penalizes ride-through violations but provides no corrective incentive within the admissible region.

\item \textbf{\emph{Duration-only}}: A duration-based variant that minimizes the accumulated excess violation duration, $\sum_{d\in\mathcal D}\sum_{k=1}^{K}[\tau_{d,k}-\tau_d^{\mathrm{trip}}]_+$, where $[x]_+=\max\{x,0\}$ and $\tau_d^{\mathrm{trip}}$ denotes the assigned trip delay of data center $d$. Penalties are applied only when consecutive violations exceed the trip delays.

\item \textbf{\emph{Ours}}: The proposed gated voltage-control objective $\ell_p(v,\tau)=(v-1)^2g_p(v,\tau)$ with $p=4$.
\end{itemize}

For a fair comparison, all methods are evaluated under identical data-center and control-device placements, demand profiles, fault conditions, initial operating states, and transfer rules. All controlled methods are subject to the same reactive-power capacity, ramp-rate, and support-budget constraints. Network operating limits, including bus-voltage bounds and branch thermal ratings, are set consistently through the ANDES case-file parameters. Simulations are conducted in \texttt{Python~3.12} using \texttt{ANDES~v2.0.0} on Google Colab equipped with an NVIDIA A100 GPU.

We evaluate the controllers using four metrics that characterize data-center retention, grid-demand loss, reactive-power utilization, and computational efficiency. Let $\chi_{d,k}^{(m)}\in\{0,1\}$ indicate whether data center $d$ has transferred from the grid by time step $k$ under method $m$, and let $K=T_{\mathrm{sim}}/\Delta t$. The metrics are defined as follows. For each method, the first three metrics are averaged across randomized trials.
\begin{itemize}[leftmargin=*]
    \item \emph{Transfer rate:}
    $
    R_{\mathrm{disc}}^{(m)}
    =\frac{1}{|\mathcal D|}
    \sum_{d\in\mathcal D}\chi_{d,K}^{(m)},
    $
    measuring the fraction of data centers transferred by the end of the simulation.

    \item \emph{Grid-demand loss:}
    $
    E_{\mathrm{loss}}^{(m)}
    =\Delta t\sum_{k=1}^{K}\sum_{d\in\mathcal D}
    \chi_{d,k}^{(m)}P_{d,k}^{\mathrm{dc}},
    $
    where $P_{d,k}^{\mathrm{dc}}$ denotes the grid-supplied demand of data center $d$ before transfer, measuring the cumulative energy demand removed from the studied grid following data-center transfers, without implying interruption of computing services.

    \item \emph{Reactive-power usage:}
    $
    E_Q^{(m)}
    =\Delta t\sum_{k=1}^{K}\sum_{j\in\mathcal C}
    |\Delta q_{j,k}^{a,(m)}|,
    $
    measuring the cumulative realized reactive-power adjustments required by each controller.

    \item \emph{Computation time:} For each trial, the total end-to-end controller computation time over $T_{\mathrm{sim}}=500$~ms is divided by $T_{\mathrm{sim}}/\Delta t=500$ control steps to obtain the average per-step computation time. We report the 95th percentile of these averages across trials to show upper-tail computational cost while reducing sensitivity to extreme outliers.
\end{itemize}

\subsection{Voltage Ride-Through Under a Limited Control Budget}
\label{subsec:main_14bus}

We first examine how different control objectives allocate limited reactive-power support to preserve data-center grid connections. 
For an illustrative trial on the IEEE 14-bus system, two data centers and their colocated control devices are located at buses~9 and~12. All control methods share a cumulative support budget of $20$~\unitBQ\ over the $500$-ms simulation, allowing direct comparison of their ride-through performance and reactive-power allocation.

Fig.~\ref{fig:nr_main_comparison} compares the PCC-voltage trajectories and reactive-power commands. Without control, both data centers transfer from the grid during the fault. Volt/VAR also incurs two transfers despite exhausting the support budget, while Quadratic reduces the outcome to one transfer with the same utilization. Both methods prioritize conventional voltage regulation without accounting for facility-specific ride-through urgency. In particular, their continued reactive-power adjustments during post-fault recovery consume support that could otherwise be reserved for critical voltage excursions. The ride-through-aware baselines improve resource utilization but still incur one transfer each: Envelope-constrained uses $14.61$~\unitBQ, whereas Duration-only uses $10.79$~\unitBQ. In contrast, the proposed method prevents both transfers using $19.00$~\unitBQ\ ($95\%$ of the budget), concentrating support during the fault and withdrawing it shortly after clearance. These results indicate that effective protection depends not simply on the amount of reactive support, but on its allocation to voltage conditions that threaten data-center ride-through.

\begin{figure}[h]
\centering
\includegraphics[width=1\linewidth]{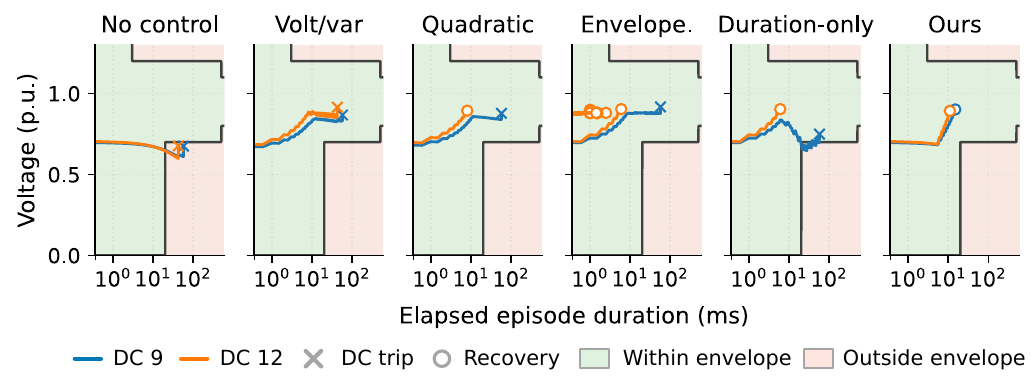}
\caption{Data-center voltage excursions versus excursion duration.}
% Each excursion is plotted against its elapsed duration. Green and red shading indicate regions within and outside the prescribed ITIC envelope, respectively. Open circles mark recovery events, and crosses mark data-center trips.}
\label{fig:voltage_duration_14bus}
\label{fig:nr_duration_grid}
\end{figure}

To further examine these differences, Fig.~\ref{fig:nr_duration_grid} shows how voltage excursions develop over time. A voltage excursion begins when a data center's PCC voltage falls below $\underline v_d$ or rises above $\overline v_d$. The x-axis shows how long each excursion has lasted, allowing us to examine both its voltage magnitude and duration. Each trajectory in Fig.~\ref{fig:nr_duration_grid} ends with either a cross, indicating that the data center disconnects, or an open circle, indicating that its voltage recovers to the normal range. An ITIC-inspired envelope is also included as a reference for assessing the severity of these excursions, rather than as the actual disconnection criterion, which is determined by each data center's assigned voltage thresholds and trip delay.

The results in Fig.~\ref{fig:nr_duration_grid} reveal distinct control behaviors. Volt/VAR and Quadratic improve voltage magnitudes but do not explicitly prioritize excursions according to their duration-dependent ride-through risks. Envelope-constrained exhibits repeated short excursions and recoveries, consistent with its violation-only objective, which provides no further corrective incentive once voltage returns within the prescribed envelope. This may leave insufficient voltage margin against subsequent excursions during a sustained fault. Duration-only accounts for excursion persistence but neglects voltage severity, leaving one data center exposed to prolonged undervoltage. In contrast, the proposed objective provides continuous corrective guidance within the envelope, with increasing sensitivity near its boundaries, enabling earlier intervention against critical excursions. It shows the benefit of prioritizing control support according to both voltage severity and excursion duration rather than nominal-voltage deviations or envelope violations alone.

\subsection{Statistical Performance Across Randomized Trials}
\label{subsec:nr_statistical}

The preceding example demonstrates the control behavior under a particular realization. To assess whether these benefits persist across different network conditions, we conduct 100 randomized trials. Table~\ref{tab:nr_main_results} summarizes the mean and standard deviation (std) of the four performance metrics.

\begin{table}[h]
\centering
\caption{IEEE 14-bus performance over 100 trials, reported as mean (std). ``Envelope.'' denotes the Envelope-constrained baseline.}
\label{tab:nr_main_results}
\footnotesize
\setlength{\tabcolsep}{2pt}
\begin{tabular}{@{}lcccc@{}}
\toprule
Method
& \shortstack{$R_{\mathrm{disc}}$\\(\%)}
& \shortstack{$E_{\mathrm{loss}}$\\(MW$\cdot$s)}
& \shortstack{$E_Q$\\(\unitBQ)}
& \shortstack{$t_{95}$\\(ms)} \\
\midrule
No control
& $100.00(0.00)$ & $28.14(5.69)$ & --- & --- \\
Volt/VAR
& $32.61(38.32)$ & $10.06(11.71)$
& $18.41(2.19)$ & $0.014$ \\
Quadratic
& $44.57(43.75)$ & $13.31(13.16)$
& $19.90(0.50)$ & $0.502$ \\
Envelope.
& $23.64(19.49)$ & $5.03(5.17)$
& $12.27(6.79)$ & $0.360$ \\
Duration-only
& $30.44(30.69)$ & $8.90(9.69)$
& $13.44(4.35)$ & $0.265$ \\
Proposed
& $13.04(24.57)$ & $3.65(7.20)$
& $16.43(3.40)$ & $0.514$ \\
\bottomrule
\end{tabular}
\end{table}

The proposed method achieves the lowest mean disconnection rate of $13.04\%$, representing a $44.8\%$ reduction relative to Envelope-constrained, the strongest baseline, along with a $27.4\%$ reduction in grid-demand loss. It also requires less reactive-power support than Volt/VAR and Quadratic, while Envelope-constrained and Duration-only achieve lower utilization at the expense of more frequent disconnections. 
Meanwhile, the additional voltage--duration weighting introduces no substantial computational overhead. The proposed method requires $0.514$~ms per control step, comparable to Quadratic ($0.502$~ms).

\subsection{Sensitivity to Deployment and Operating Conditions}
\label{subsec:sensitivity}

\begin{figure}[h]
\centering
\includegraphics[width=\linewidth]{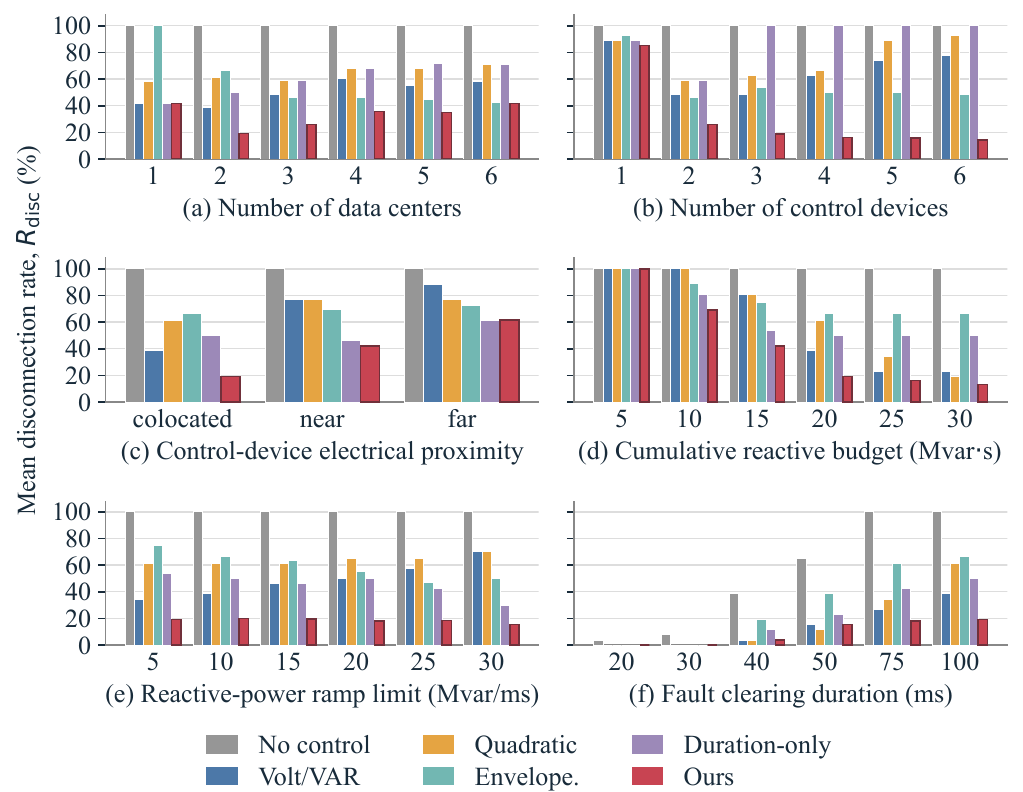}
\caption{Mean disconnection-rate sensitivity in the IEEE 14-bus system. }
\label{fig:nr_sensitivity}
\end{figure}

In practice, voltage-control effectiveness depends on data-center locations, reactive-support deployment, and operating conditions. To assess the robustness of the proposed method under these variations, Fig.~\ref{fig:nr_sensitivity} presents six sensitivity studies.

Figs.~\ref{fig:nr_sensitivity}(a)--(c) investigate the effects of data-center penetration and support-device deployment. As the number of data centers varies, the proposed method consistently achieves the lowest or comparable disconnection rates, demonstrating its effectiveness in coordinating support across multiple facilities. Increasing the number of control devices generally improves ride-through performance, with the proposed method maintaining disconnection rates near $15\%$ when sufficient devices are available. We further examine electrical proximity by placing support devices at data-center buses (``colocated'') or ranking eligible hosting buses by their electrical distances to existing data centers and selecting the $|\mathcal C|$ nearest (``near'') or farthest (``far'') buses. Although electrically distant support reduces the effectiveness of all methods, the proposed method maintains favorable performance across these deployment configurations.

Figs.~\ref{fig:nr_sensitivity}(d)--(f) evaluate the effects of reactive-power constraints and fault severity. Under restrictive cumulative budgets, all methods experience increased disconnections due to limited corrective capability. As more support becomes available, the proposed method achieves substantially lower disconnection rates, indicating more effective utilization of reactive-power resources. Its performance also remains relatively stable across different ramp-rate limits. Finally, longer fault-clearing durations increase the difficulty of maintaining ride-through, yet the proposed method limits the disconnection rate to approximately $20\%$ even under a $100$-ms fault.

Overall, these results demonstrate that the proposed voltage--duration objective improves data-center ride-through across diverse deployment configurations, resource constraints, and disturbance conditions, supporting its applicability to practical networks with heterogeneous data-center integration.

\subsection{Property-by-Property Ablation Study}
\label{subsec:nr_ablation}

To isolate the contribution of each proposed property, we construct four ablation objectives, each deliberately violating one of P1--P4 while retaining the other three. All variants are evaluated under identical conditions. 
As in Section \ref{subsec:objective_family}, our proposed objective is $\ell_4(v,\tau)=\varepsilon z^4$, where $z=|v-1|/r_\pm(\tau)$ denotes the normalized voltage deviation. The ablation objectives are constructed as
\begin{subequations}\label{eq:ablation_objectives}
\begin{align}
\ell_{\neg P1}(v,\tau)
&=\varepsilon\left(\frac{|v-1|}{r_\pm(\tau)+\xi}\right)^4,
\\
\ell_{\neg P2}(v,\tau)
&=\varepsilon\left[(1-\eta\cdot \tau/T_{\mathrm{sim}})z^4+\eta \cdot \tau/T_{\mathrm{sim}}z^2\right],
\\
\ell_{\neg P3}(v,\tau)
&=\varepsilon\left(\frac{[z-\zeta]_+}{1-\zeta}\right)^4,
\\
\ell_{\neg P4}(v,\tau)
&=\varepsilon z,
\end{align}
\end{subequations}
where $\xi>0$, $0<\eta\leq1$, and $0<\zeta<1$. 

\begin{figure}[h]
    \centering
    \includegraphics[width=\linewidth]{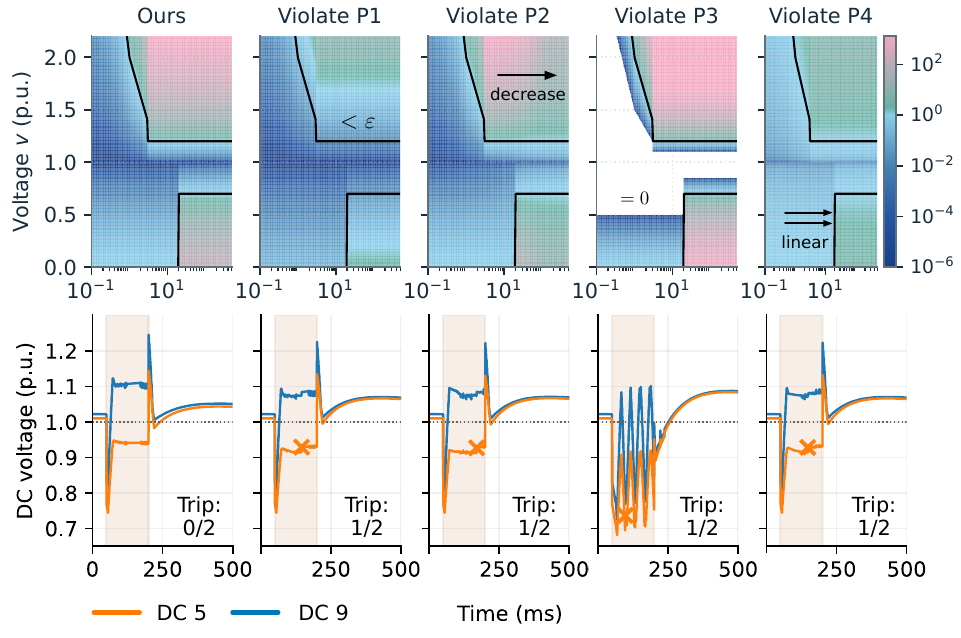}
    \caption{Property-by-property ablation of the proposed objective. Upper panels show the original and modified loss surfaces, with black curves indicating the prescribed voltage--duration envelope. Lower panels show the corresponding PCC voltage responses and data-center trips under identical conditions. Parameters: $\xi=0.5$, $\eta=0.9$, and $\zeta=0.3$.}
    \label{fig:nr_ablation}
\end{figure}

The upper panels of Fig.~\ref{fig:nr_ablation} illustrate the distinct effects of these modifications. Removing P1 enlarges the effective voltage margins, assigning insufficient penalties to deviations beyond the prescribed envelope. Removing P2 reverses the intended temporal priority, reducing the penalty as an out-of-envelope violation persists. Removing P3 introduces a zero-loss region around nominal voltage, eliminating corrective gradients for $z\leq\zeta$. Removing P4 replaces the increasing marginal penalty of the objective with constant voltage sensitivity, weakening the emphasis on deviations approaching the envelope boundaries. Each variant also preserves the other three properties, allowing their individual roles to be examined.

The lower panels of Fig.~\ref{fig:nr_ablation} reveal the corresponding control consequences. While the proposed objective retains both data centers, each ablation results in one transfer under the same disturbance. The P1 ablation misaligns corrective priorities with the prescribed voltage margins, whereas the P2 ablation weakens the response to sustained violations. Without P3, the loss of corrective guidance near nominal voltage is accompanied by pronounced voltage oscillations. Without P4, insufficient emphasis on near-boundary deviations leaves one data center unprotected. Together, these observations highlight the complementary roles of the four properties: P1 establishes envelope consistency, P2 accounts for violation persistence, P3 maintains corrective direction, and P4 strengthens near-boundary intervention. Their combination produces more effective ride-through protection than any of the examined incomplete constructions.

\subsection{Scalability to Larger Transmission Networks}
\label{subsec:nr_scalability}

To assess scalability beyond the IEEE 14-bus system, we extend the evaluation to the New England 39-bus and NPCC 140-bus transmission networks, each with 100 trials. As shown in Table~\ref{tab:nr_scalability}, the proposed method maintains the lowest disconnection rates of $18.75\%$ and $27.40\%$, representing reductions of $45.5\%$ and $27.5\%$ relative to Envelope-constrained (strongest baseline), respectively. Despite higher absolute disconnection rates than on the IEEE 14-bus system, the proposed method consistently outperforms competing approaches in both data-center retention and grid-demand loss, demonstrating its effectiveness across different network topologies and sizes.

\begin{table}[h]
\centering
\caption{Scalability over 100 paired trials.
Pairs list New England 39-bus/NPCC 140-bus results.
}
\label{tab:nr_scalability}
\footnotesize
\setlength{\tabcolsep}{3pt}
\renewcommand{\arraystretch}{1.12}
\begin{tabular}{@{}lcccc@{}}
\toprule
Method
& \shortstack{$R_{\mathrm{disc}}$\\(\%)}
& \shortstack{$E_{\mathrm{loss}}$\\(MW$\cdot$s)}
& \shortstack{$E_Q$\\(\unitBQ)}
& \shortstack{$t_{95}$\\(ms)} \\
\midrule
No control
& $46.88/57.53$ & $67.32/152.84$ & --- & --- \\
Volt/VAR
& $34.38/49.60$ & $30.46/78.52$
& $49.84/48.76$ & $0.046/0.047$ \\
Quadratic
& $37.50/53.20$ & $34.91/85.64$
& $59.98/59.63$ & $0.767/0.894$ \\
Envelope.
& $34.38/37.80$ & $20.93/58.27$
& $25.57/28.76$ & $0.600/0.832$ \\
Duration-only
& $34.38/44.20$ & $26.63/69.31$
& $34.92/40.12$ & $0.441/0.672$ \\
Proposed
& $18.75/27.40$ & $12.41/39.82$
& $44.29/46.18$ & $0.802/0.921$ \\
\bottomrule
\end{tabular}
\end{table}

The per-step computation time increases from $0.514$~ms on the 14-bus system to $0.802$~ms and $0.921$~ms on the larger networks, remaining comparable to the optimization-based baselines. These measurements are obtained in a general-purpose computing environment, with further acceleration potentially achievable through dedicated hardware and implementation optimization. Overall, the results demonstrate favorable scalability in both ride-through performance and computational efficiency.

\section{Conclusion}
\label{sec:conclusion}

This paper proposes a property-driven voltage-control framework to reduce fault-induced transfers of large data-center loads. Using prescribed voltage--duration ride-through envelopes, we establish four properties that characterize ride-through consistency and corrective behavior. We then derive conditions under which gated quadratic objectives satisfy these properties and construct a closed-form, envelope-normalized objective. 
Numerical studies demonstrate improved data-center retention under limited reactive-power support. On the IEEE 14-bus system, the proposed method reduces the mean disconnection rate by $44.8\%$ and grid-demand loss by $27.4\%$ relative to the strongest baseline. These results demonstrate the value of incorporating voltage--duration requirements directly into the control objective, rather than relying solely on nominal-voltage tracking or envelope-violation penalties.

\bibliographystyle{IEEEtran}
\bibliography{ref}

\end{document}